\documentclass[12pt,a4paper,twoside]{article}

\usepackage{graphicx}
\usepackage{amsmath}

\newcommand{\bd} {\ensuremath{\mathrm{B^{0}}}}
\newcommand{\kstar} {\ensuremath{\mathrm{K^{*0}}}}
\newcommand{\qs} {\ensuremath{q^{2}}}
\newcommand{\CP} {\ensuremath{CP}}

\newcommand{\bs}              {\ensuremath{\mathrm{B^{0}_{s}}}}

\newcommand{\dgs}              {\ensuremath{\mathrm{\Delta\Gamma_{s}}}}
\newcommand{\phis}              {\ensuremath{\mathrm{\Phi_{s}}}}
\newcommand{\betas}              {\ensuremath{\mathrm{\beta_{s}}}}

\newcommand{\fl}              {\ensuremath{\mathrm{F_{L}}}}
\newcommand{\afb}              {\ensuremath{\mathrm{A_{FB}}}}

\newcommand{\jpsi}              {\ensuremath{\mathrm{\mathrm{J}/\psi}}}
\newcommand{\jpsitwos}              {\ensuremath{\mathrm{\psi(2S)}}}

\newcommand{\bsjpsiphi}               {\ensuremath{\mathrm{B^{0}_{s}}\to \mathrm{J}/\psi \phi}}

\newcommand{\bjpsik}               {\ensuremath{\mathrm{B^{\pm}}\to \mathrm{J}/\psi \mathrm{K}^{\pm}}}

\newcommand{\bjpsikp}               {\ensuremath{\mathrm{B^{+}}\to \mathrm{J}/\psi \mathrm{K}^{+}}}
\newcommand{\bjpsikm}               {\ensuremath{\mathrm{B^{-}}\to \mathrm{J}/\psi \mathrm{K}^{-}}}

\newcommand{\bkstarmumu}               {\ensuremath{\mathrm{B^{0}}\to \mathrm{K}^{*0} \mu^{+} \mu^{-}}}
\newcommand{\bkmumu}               {\ensuremath{\mathrm{B^{\pm}\to \mathrm{K}^{\pm} \mu^{+} \mu^{-}}}}

\newcommand{\bquark} {\ensuremath{\mathrm{\overset{\textbf{\fontsize{4pt}{4pt}\selectfont(---)}}{b}}}-quark}

\title{Angular analysis of the decay 
$\bf  \bkstarmumu$
and studies  of mixing and ${\boldmath {CP}}$ violation 
in the $\bf B_S^0$ system with the ATLAS detector}

\author{Jochen Schieck on behalf of the ATLAS Collaboration \\
Ludwig-Maximilians-Universit\"at M\"unchen, \\ Am Coulombwall 1, 85748 Garching, Germany and \\
  Excellence Cluster Universe, Boltzmannstr. 2, \\ 85748 Garching, Germany. \\       
          E-mail: {Jochen.Schieck@lmu.de}}

\begin{document}
\maketitle
\abstract{The measurement of B meson decay properties provides an alternative approach for searches for physics beyond the
Standard Model. We present two different measurements performed with data taken in 2011 with the ATLAS detector  at the 
Large Hadron collider. The first measurement
is based on an angular analysis of the decay  \bkstarmumu, and in the second measurement the decay width difference \dgs\ and the weak 
phase \phis\ of the \bs\ system is determined using the decay \bsjpsiphi. For the second measurement a probability for the flavour quantum number
of the \bquark\ at production time is determined.}

\section{Introduction}
With the discovery of the Higgs boson the Standard Model of Particle Physics is complete with all expected particles being 
observed. However, we know
that the Standard Model is not enough to describe all phenomena of nature. The most striking example might be the 
absence of a description of gravity. Theses shortcomings clearly point to the existence of, until now 
undiscovered, physics beyond the Standard Model. At the Large Hadron Collider (LHC) at CERN two orthogonal approaches are 
currently being pursued. Particles from physics beyond the Standard Model can be produced at processes 
with large momentum transfer and subsequently observed directly in the detector. 
Alternatively particles from physics  beyond the Standard Model can shortly appear in quantum loops and shape 
the outcome of the process without 
being directly observed. In physics with B mesons involved rare processes, together with a very precise prediction from Standard Model 
calculations, offer such an alternative search path.  \par
In this paper we present the analysis of two different B meson decays, based on data taken with the ATLAS detector~\cite{Aad:2008zzm} 
at the LHC in the year 2011. First an angular analysis of the decay \bkstarmumu\ is discussed and a measurement of
 the longitudinal polarisation \fl\ and the forward-backward asymmetry \afb\ is presented~\cite{bibkstar}. This analysis is 
followed by a measurement of the decay width difference and the weak phase of the \bs\ using the 
decay \bsjpsiphi~\cite{bibjpsiphi}.
\section{The ATLAS detector and the data sample}
The measurements presented in this paper are performed with data taken with the ATLAS detector~\cite{Aad:2008zzm}
in 2011. Overall $4.9\,  \mathrm{fb}^\mathrm{-1}$ of data are collected at a centre-of-mass energy of 7 TeV 
with a maximum luminosity of about $3.5\times10^{33}\, \mathrm{cm}^{-2}$s$^{-1}$. The main trigger used to select events 
of interest is based on the observation of two muons in the final state, with the invariant mass of the di-muon pair 
being within a certain mass range.
\subsection{Detector Performance}
\begin{figure}[]
\begin{center}
\includegraphics[width=.425\textwidth]{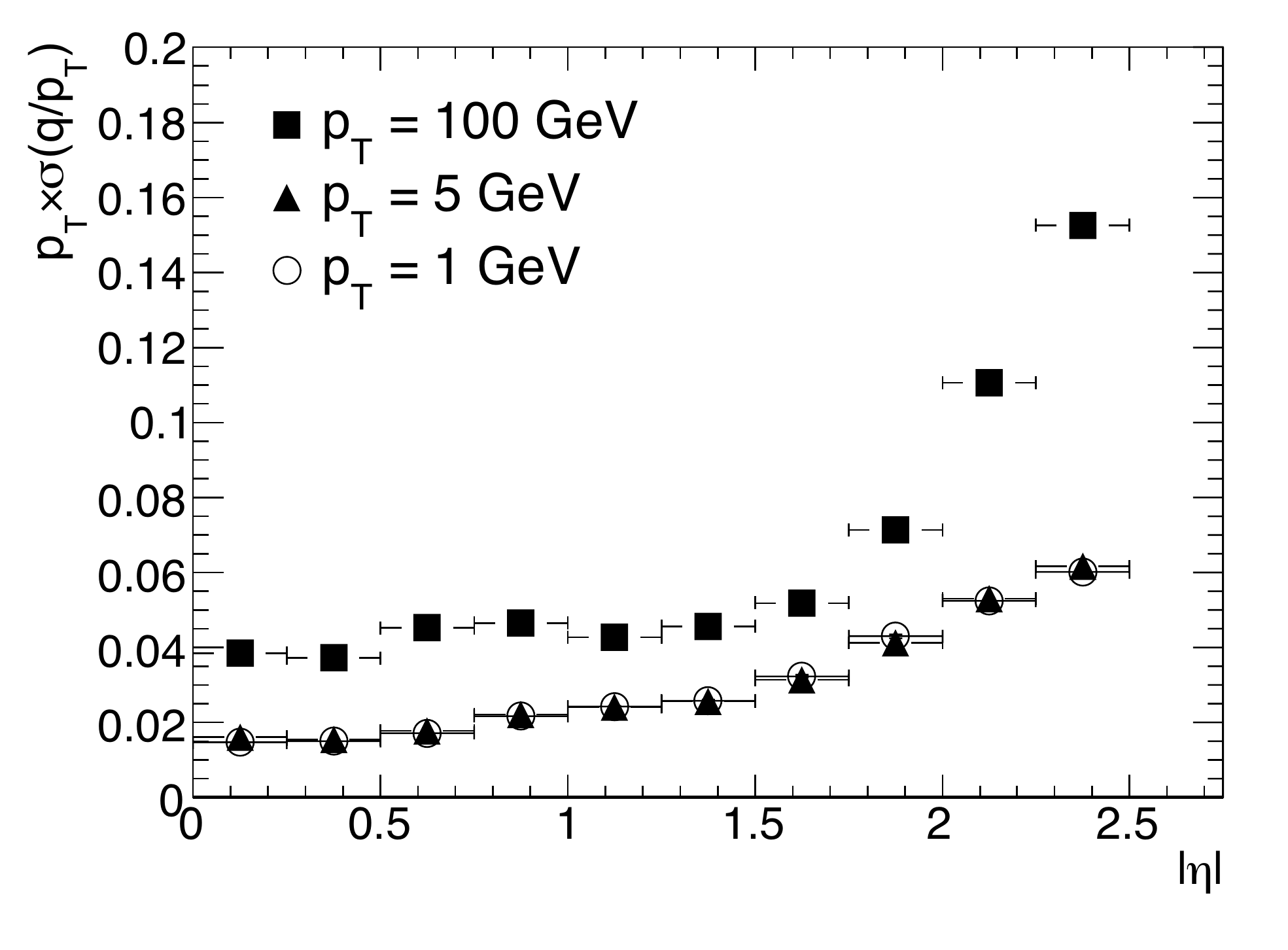}
\includegraphics[width=.465\textwidth]{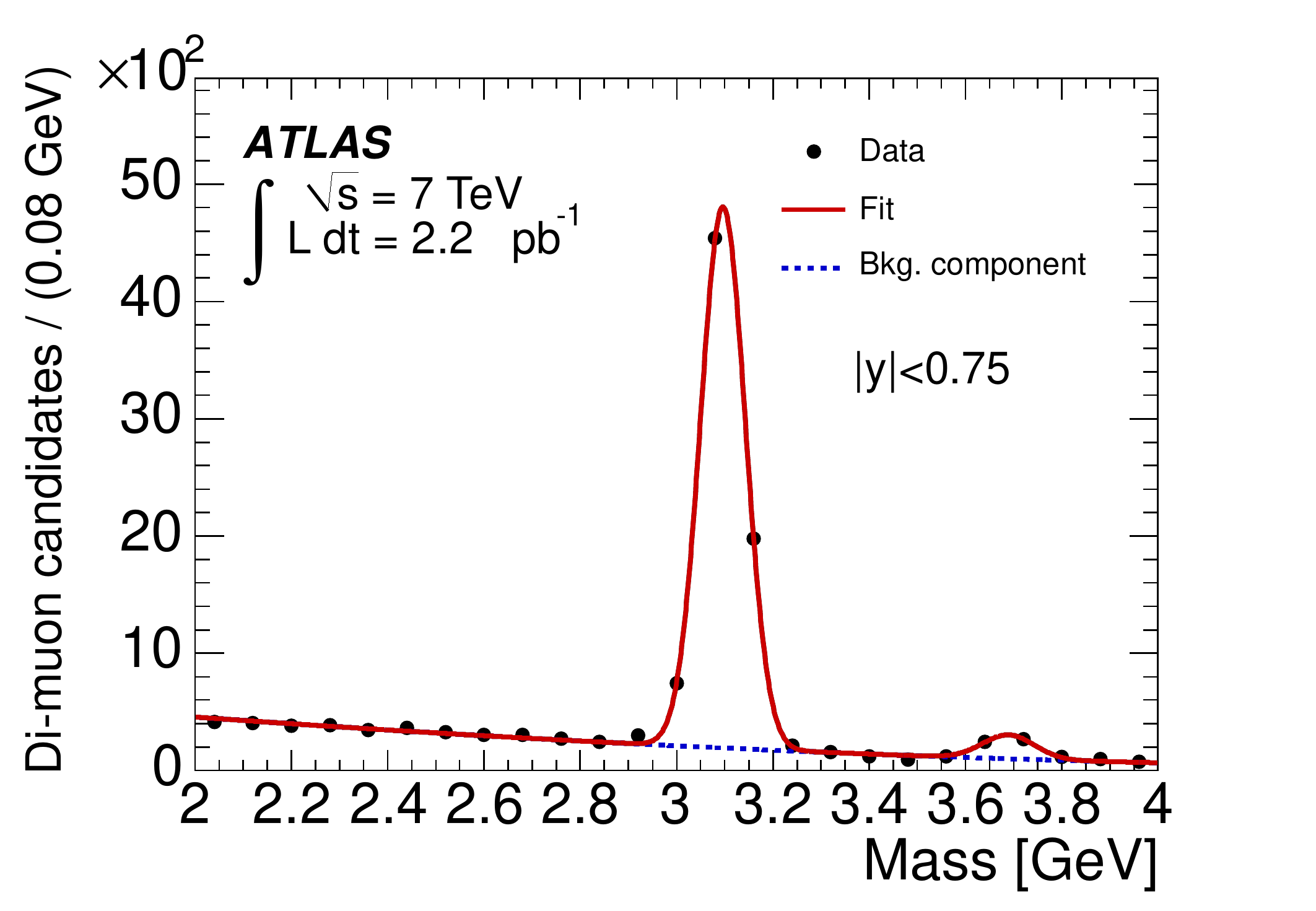}
\caption{Impact parameter resolution as a function of pseudo rapidity $\eta$ for charged particles with a momentum of 1, 5 
and 10 GeV (left)~\cite{Aad:2008zzm}. Invariant mass of selected \jpsi\ reconstructed in the central barrel region 
of the detector (right)~\cite{Aad:2011sp}.}
\label{fig1}
\end{center}
\end{figure}

The measurements performed for these analyses are mainly based on information taken with the inner detector (ID), the main tracking device 
of the detector, and the muon spectrometer (MS), the sub-detector responsible for identifying muons in the event. The ID is 
used to reconstruct the trajectories of charged particles in order to determine their momentum and their impact parameter with respect to
the interaction point in the plane transverse to the beam line. The
measurement of the impact parameter is of interest in order to reconstruct the lifetime of weakly decaying B hadrons. The 
expected impact parameter resolution, determined with simulated events, is shown in Figure~\ref{fig1}. The momentum 
measurement is needed to determine the invariant mass of a decaying hadron in order to separate signal from background events.
As an example the mass resolution of a \jpsi\ is shown in Figure~\ref{fig1}. The mass resolution obtained for \jpsi\ 
reconstructed in the central barrel region is $\sigma_{\mathrm{m}}=46\pm1\,$MeV. 
The ATLAS detector contains no dedicated sub-detector to distinguish between charged kaons and pions for most momenta, leading to 
a worse signal to noise ratio for fully reconstructed B meson decays, compared to experiments with dedicated particle identification. 
\section{Measurement of the angular distribution in the decay \bkstarmumu}
\begin{figure}
\begin{center}
\includegraphics[width=.50\textwidth]{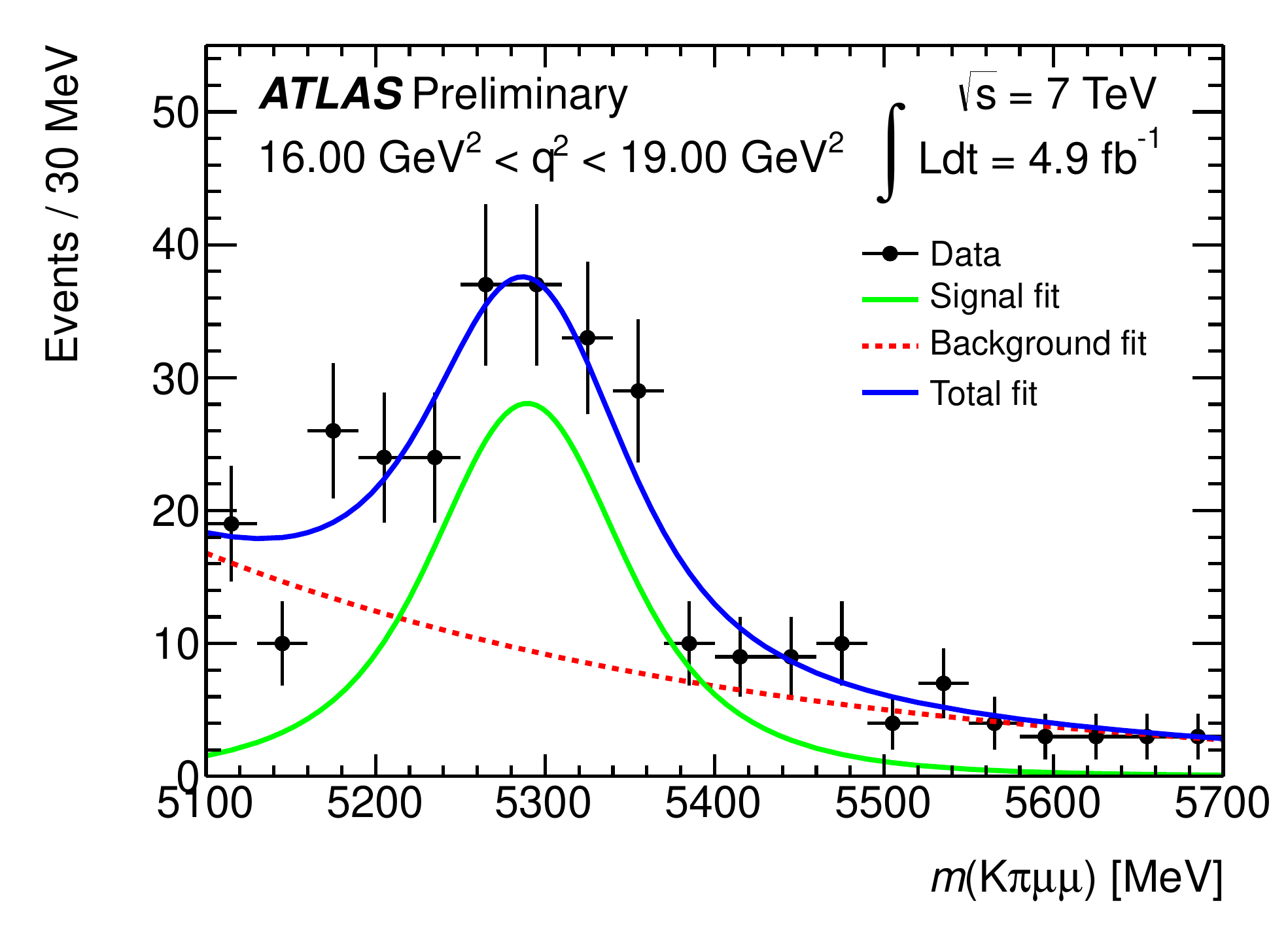}
\includegraphics[width=.49\textwidth]{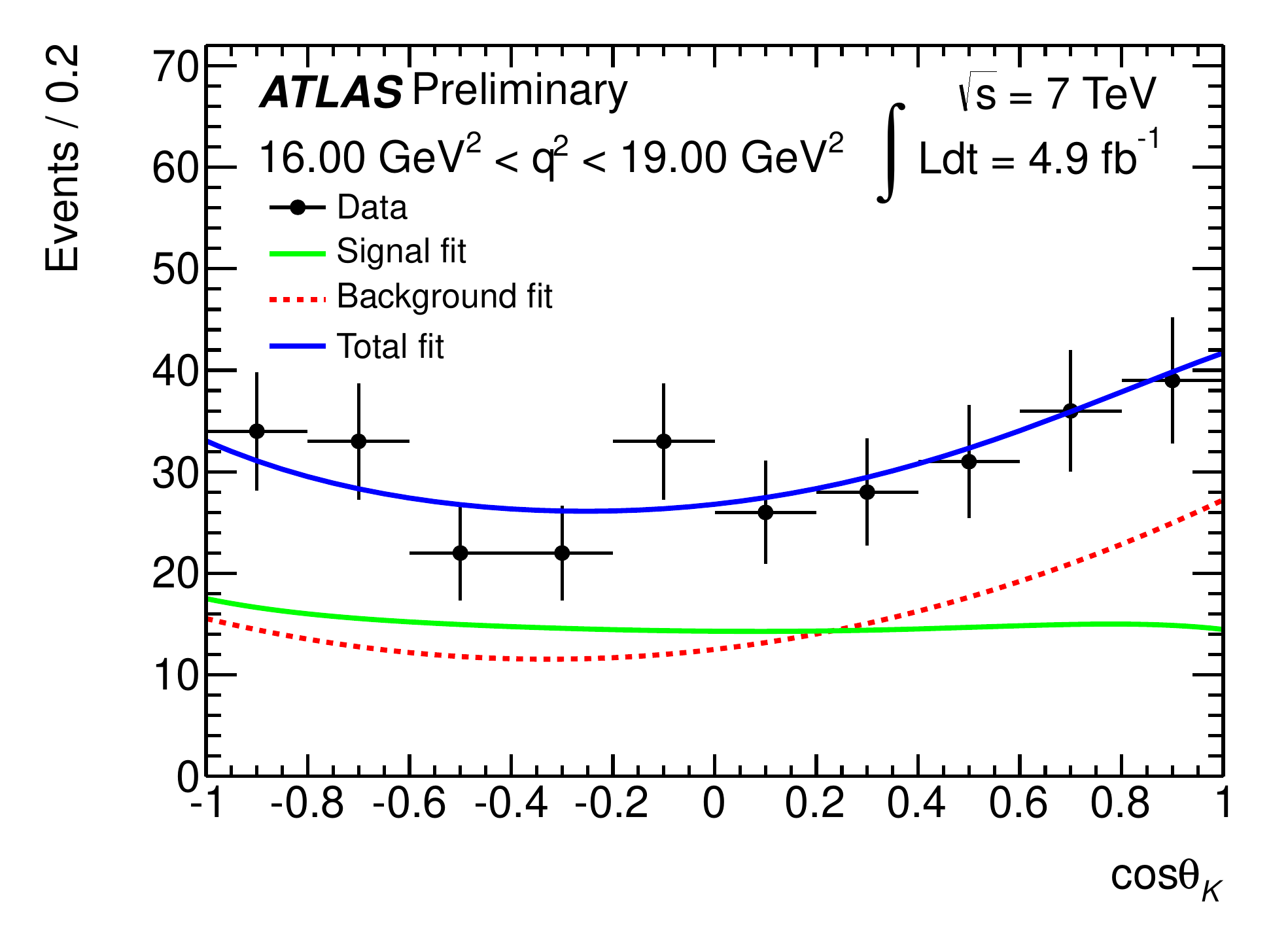}
\caption{Invariant mass distribution (left)~\cite{bibkstar} and $\theta_{K}$ distribution (right)~\cite{bibkstar} of the selected \bkstarmumu\ events in the \qs\ region between 16 GeV$^2$ $<$ \qs\ $<$ 19 GeV$^2$.}
\label{fig2}
\end{center}
\end{figure}

The semi-rare decay of a \bd\ into a \kstar\ and a muon pair takes place via a loop diagram, including a second order transition 
from a b-quark to a s-quark. The decay can be described by four kinematic variables, \qs, the invariant mass of the di-muon pair and 
three angles of the four-body final state. Contributions from physics beyond the Standard Model could alter the decay amplitude as well as the 
distribution of the kinematic variables. A full description of the analysis summarised here can be found at~\cite{bibkstar}. 
The branching ratio of this decay channel is in the order of $10^{-6}$ and, in the invariant mass region 
between $4900\,$MeV $<$ m(K$\pi\mu\mu$) $<$ $5700\,$MeV, the estimated number of signal events amounts to $466\pm34$ with 
$1132\pm43$ background events. 
Events are selected with a cut based selection procedure, optimised on simulated events. Events with a \qs\ value consistent with a 
muon pair from a \jpsi\ or a \jpsitwos\ decay are excluded. 
The data set is subdivided in five different \qs\ regions (see Table~\ref{tabresult}) and the relative small signal yield for each of these regions does not allow a 
full angular analysis using all three decay angles. For this reason two out of the three angles are integrated out.
In this analysis two differential decay distributions are used: the differential decay rate with respect to the angle $\theta_{L}$, the angle between between the $\mu^{+}$ and the and the opposite direction of the \bd\ in the di-muon rest frame and the differential decay rate with respect to the angle $\theta_{K}$, the angle between the $K^{+}$ and the opposite direction of the \bd\ in the \kstar\ rest frame. 
These differential decay rates are then used to determine, with an unbinned maximum likelihood fit, the longitudinal polarisation of 
the \kstar, \fl, and the forward backward asymmetry of the di-muon system, \afb. A sequential approach is used, where first the
number of signal and background events is determined with a fit to the invariant mass distribution, followed by a fit to the angular 
distributions, with the number of signal and background events being fixed to the values obtained from the fit to the invariant mass.
The measured angular decay distribution is corrected for acceptance effects using Monte Carlo events with the response of the detector
fully simulated.
Possible distortions originating from interchanging the kaon and the pion in the \kstar\ decay are taken into account. The invariant mass
 and the $\theta_{K}$ distribution for events with a \qs\ value between $16\,$GeV$^2$ and $19\,$GeV$^2$
are shown in Figure~\ref{fig2}. \par
Several sources of systematic uncertainties are studied. The sequential fit approach only introduces a significant uncertainty for 
very small event samples. In the \qs\ region between $2\,$GeV$^{2}$ and $4.3\,$GeV$^{2}$ only $19\pm8$ events are selected, leading 
to  the sequential fit procedure being the dominant systematic uncertainty. In the other \qs\ regions 
misidentified \bkmumu\ events tend to dominate
the \afb\ measurement and the acceptance correction procedure the \fl\ measurement. The overall uncertainty is dominated by the
statistical uncertainty.  A summary of the number of signal events together with
the longitudinal polarisation and the forward backward asymmetry for the different \qs\ regions is presented in Table~\ref{tabresult}. \par
\begin{table}
\begin{center}
\begin{small}
  \renewcommand{\arraystretch}{1}
  \begin{tabular}{ c   c   c   c  }
    \hline
    \qs\ range (GeV$^{2}$) & $\mathrm{N_{sig}}$ & $\afb$ & $\fl$  \\ \hline
    $~ ~ 2.00 < q^2 < ~ ~4.30$ & $~ ~ 19 \pm ~ ~ 8$ &
        $0.22\pm 0.28\pm0.14$ & $0.26\pm0.18\pm0.06$  \\
    $~ ~ 4.30 < q^2 < ~ ~8.68$ & $~ ~ 88 \pm 17$ &
        $0.24\pm0.13\pm0.01$ & $0.37\pm0.11\pm0.02$  \\
    $10.09 < q^2 < 12.86$ & $138 \pm 31$ &
        $0.09\pm0.09\pm0.03$ & $0.50\pm0.09\pm0.04$  \\
    $14.18 < q^2 < 16.00$ & $~ ~ 32 \pm 14$ &
        $0.48\pm0.19\pm0.05$ & $0.28\pm0.16\pm0.03$  \\
    $16.00 < q^2 < 19.00$ & $149 \pm 24$ &
        $0.16\pm0.10\pm0.03$ & $0.35\pm0.08\pm0.02$  \\ \hline
  \end{tabular}
\end{small}
\end{center}
\caption{Fit result including statistical and systematic uncertainty for the number of signal events, $\mathrm{N_{sig}}$, 
the longitudinal polarisation, \fl, and the forward backward asymmetry, \afb, obtained in the the different regions \qs~\cite{bibkstar}. }
  \label{tabresult}
\end{table}
\begin{figure}
\begin{center}
\includegraphics[width=.45\textwidth]{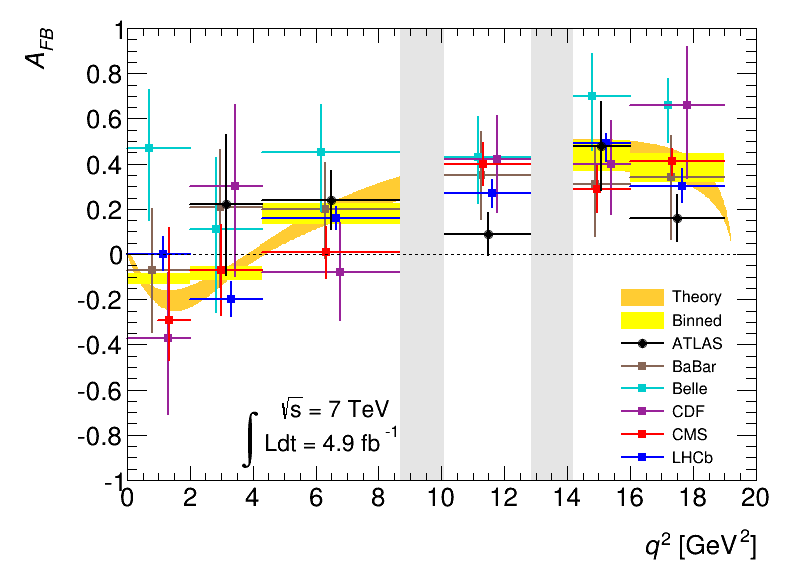}
\includegraphics[width=.45\textwidth]{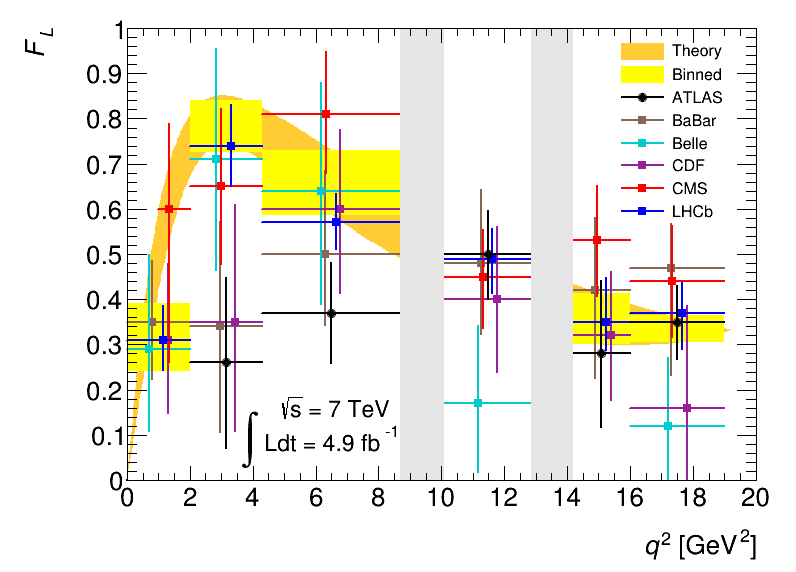}
\caption{Comparison of the measurement of \afb\ and \fl\ as a function of \qs\ obtained by different experiments. The theoretical
prediction is shown as well~\cite{plotkstar}.}
\label{fig3}
\end{center}
\end{figure}
A graphical presentation of this measurement, together with results obtained by other collaborations, is summarised 
in Figure~\ref{fig3}. In general,
all measurements are consistent with the theoretical prediction obtained from the Standard Model. 
The ATLAS results are most competitive in the high \qs\ region, both for \afb\ and \fl. 
Using the data sample already collected with the ATLAS experiment in 2012 and applying
an optimised selection procedure with dedicated cuts in the high \qs\ region, data taken with the ATLAS experiment have the opportunity
to contribute significantly to this measurement. We also expect an enhanced number of selected events due to an improved
triggering and delayed processing scheme in 2012.
\section{Measurement of \dgs\ and \phis\ in the decay \bsjpsiphi}
Contributions from physics beyond the Standard Model can also modify the \bsjpsiphi\ decay parameters and in particular the \CP\ violating phase \phis.
In this decay \CP\ violation occurs in the interference of neutral \bs\ mixing and the subsequent decay. The Standard Model predicts very small \CP\
violating effects in this decay and any measurement of significant \CP\ violation would be a clear sign for physics beyond the Standard Model. With a time
dependent angular analysis of the decay \bsjpsiphi\ the angle \betas\ of the \bs\ unitarity triangle can be measured, with 
$-2\betas$ almost being identical to the \CP\ violating phase \phis.
A measurement of \phis\ and the decay width difference \dgs, without identifying the flavour quantum number of the \bquark\ at 
production time (untagged), was published recently in~\cite{Aad:2012kba}. Here we present an update
of this measurement, including an additional identification  of the the \bquark\ flavour quantum number at production. A detailed summary of this analysis can
be found at~\cite{bibjpsiphi}.
\subsection{Determination of the \bquark\ quantum flavour number at production time}
\begin{figure}
\begin{center}
\includegraphics[width=.52\textwidth]{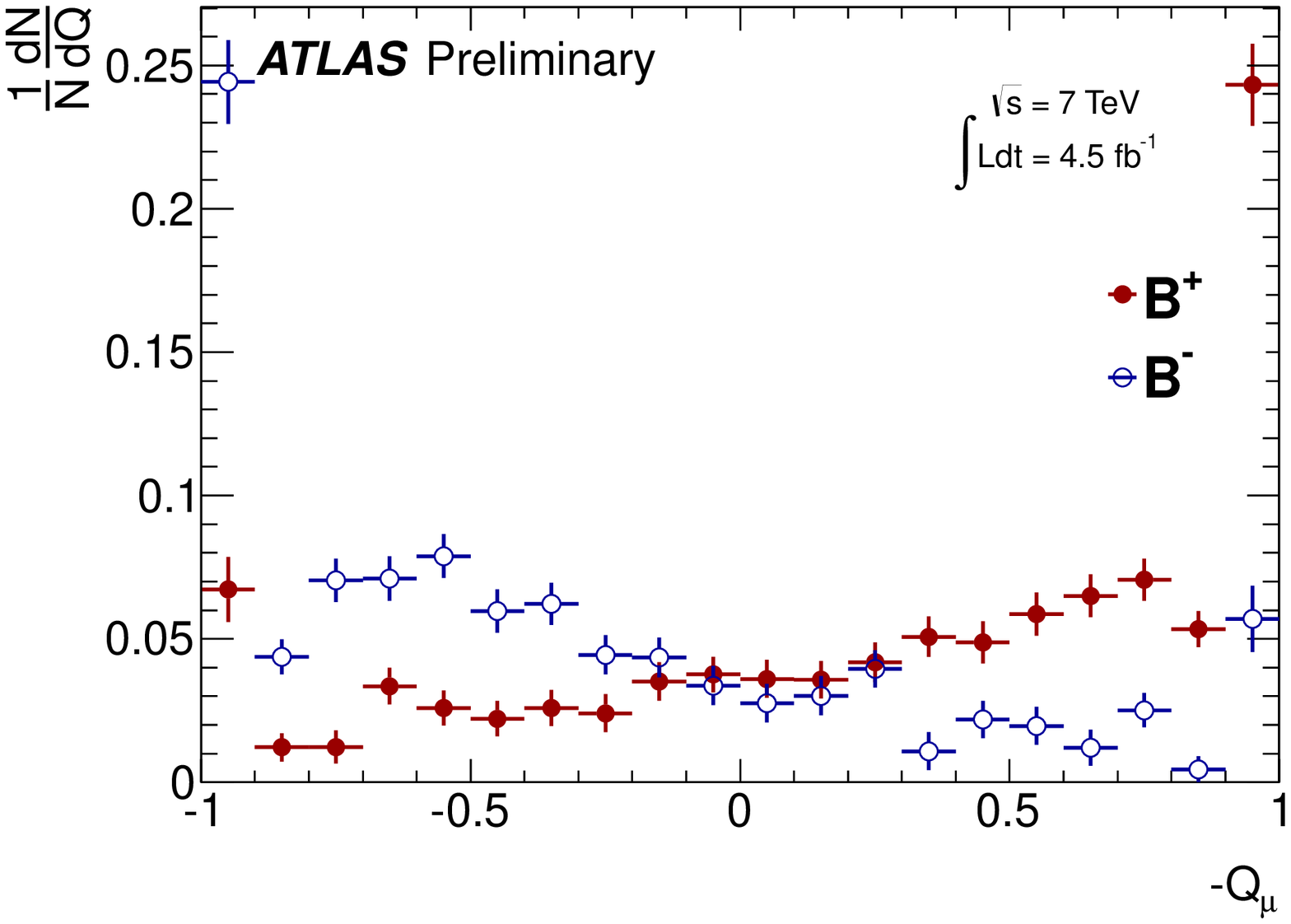}
\includegraphics[width=.38\textwidth]{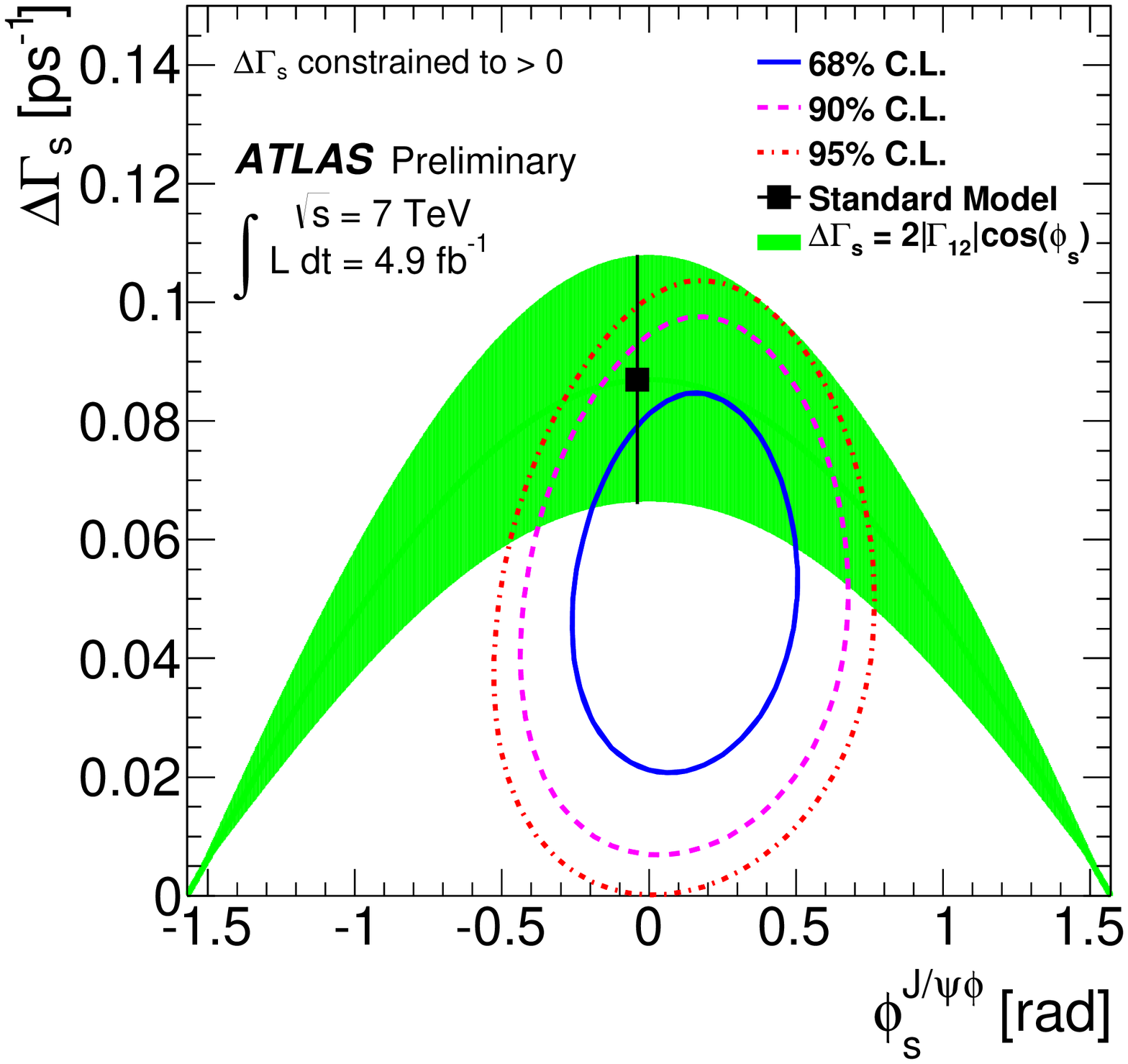}
\caption{Distribution of the muon cone charge $Q_{\mu}$ for muon tracks reconstructed in the ID and the MS (left)~\cite{bibjpsiphi}. 
Likelihood contours for \phis\ and \dgs\ obtained from the unbinned maximum likelihood fit together with the green band being the
prediction from the Standard Model (right)~\cite{bibjpsiphi}. }
\label{fig4}
\end{center}
\end{figure}

The event selection, as well as the unbinned maximum likelihood fit to determine the decay parameters, follows closely the procedure
published in~\cite{Aad:2012kba}. For the determination of  the \bquark\ flavour quantum number of the \bs, two different tagging methods are 
used. Both methods are based on the so-called opposite tagging technique (OST), which takes advantage of the fact that \bquark s are produced
via the strong interaction and the production of a b-quark is always accompanied by the production of a $\overline{\mathrm b}$-quark. 
The tagging algorithm identifies, besides the fully reconstructed signal decay, a second \bquark\ decay. The production
of a b or $\overline{\mathrm b}$ is distinguished by 
either calculating the muon cone charge $Q_\mu$, using tracks around the lepton from the leptonic \bquark\ decay, or, in case 
of  hadronic decay, using a jet charge 
$Q_{\mathrm{jet}}$, calculated 
from tracks associated with the opposite \bquark\ decay. For calibration of the OST method the self-tagging decay \bjpsik\ is used, 
where the charge of the kaon identifies the flavour quantum number of the \bquark\ at production. As an example Figure~\ref{fig4} shows the 
muon cone charge $Q_{\mu}$ determined separately for events with a reconstructed \bjpsikp\ or \bjpsikm\ decay. The determined
value of $Q_{\mu}$ shows a clear dependence on the B meson flavour quantum number, indicating the tagging capabilities of the method.
The combined OST methods reach a tagging efficiency of $\epsilon=(32.1\pm0.1)\% $ and a dilution of $\cal{D}$$=(21.3\pm0.08)\%$, leading
to a overall tagging power of $(1.45\pm0.05)\%$, with the uncertainty reflecting the statistical uncertainty only. \par
The invariant mass and the lifetime of the reconstructed \bsjpsiphi\ candidate, together with their uncertainty,
are used in an unbinned maximum likelihood fit. The \CP\ content of the vector-vector final state is determined by
the angular distribution of the final state and is included in the fit as well. The fit is completed using the
candidate-by-candidate probability of the quark produced was a b or a  $\overline{\mathrm b}$-quark determined with OST. 
The main parameters returned from the fit are $\dgs=(0.053\pm0.021\pm0.009)\,$ps$^{-1}$ and
$\phis=0.12\pm0.25\pm0.11$. The systematic uncertainty for \dgs\ is dominated by using  an uncorrelated description of
the angles from background events in the default fit. For the weak phase \phis\ the dominant uncertainty 
is associated with the OST; that is largely statistical in nature.
The correlation between the two parameters is shown 
in Figure~\ref{fig4}. The result obtained is consistent with the prediction expected from the Standard Model. The statistical uncertainty on
the measurement of \phis\ decreased by almost $40\%$, compared to the measurement without using information of 
the \bquark\ flavour quantum number. The mean values and the uncertainty of all other observables did not change with
the use of the tagging information.



\begin{thebibliography}{99}
\bibitem{plotkstar}
The figures containing a summary of currently available measurements are provided by Patrick Jussel.
\end{thebibliography}
\end{document}